\documentclass[a4paper,aps,pra,superscriptaddress,notitlepage,nofootinbib,reprint,reqno]{revtex4-1}
             \usepackage{amssymb,amsfonts,amsmath}
             \usepackage[utf8]{inputenc}
             \usepackage[usenames, dvipsnames,svgnames,table]{xcolor}
             \usepackage[colorlinks=true, allcolors=NavyBlue, breaklinks=true]{hyperref}
\usepackage{dsfont} 

\newcommand{\id}{\mathds{1}}

\DeclareMathOperator{\tr}{tr}

\let\originalleft\left
\let\originalright\right
\renewcommand{\left}{\mathopen{}\mathclose\bgroup\originalleft}
\renewcommand{\right}{\aftergroup\egroup\originalright}
             \usepackage{siunitx}
             \usepackage{graphicx}
             \usepackage{mathrsfs}
             \usepackage{braket}
\begin{document}
\title{Causally nonseparable processes admitting a causal model}
\author{Adrien Feix}
\author{Mateus Ara{\'u}jo}
\author{{\v C}aslav Brukner}
\affiliation{Faculty of Physics, University of Vienna, Boltzmanngasse 5, 1090 Vienna, Austria}
\affiliation{Institute for Quantum Optics and Quantum Information (IQOQI), Boltzmanngasse 3, 1090 Vienna, Austria}
\date{\today}
\begin{abstract}
A recent framework of quantum theory with no global causal order predicts the existence of ``causally nonseparable'' processes. Some of these processes produce correlations incompatible with any causal order (they violate so-called ``causal inequalities'' analogous to \emph{Bell inequalities}) while others do not (they admit a ``causal model'' analogous to a \emph{local model}). Here we show for the first time that bipartite causally nonseparable processes with a causal model exist, and give evidence that they have no clear physical interpretation. We also provide an algorithm to generate processes of this kind and show that they have nonzero measure in the set of all processes. We demonstrate the existence of processes which stop violating causal inequalities but are still causally nonseparable when mixed with a certain amount of ``white noise''. This is reminiscent of the behavior of Werner states in the context of entanglement and nonlocality. Finally, we provide numerical evidence for the existence of causally nonseparable processes which have a causal model even when extended with an entangled state shared among the parties.
\end{abstract}
\maketitle
\section{Introduction}
\label{sec:introduction}
It is well-known that quantum mechanics is at odds with naive notions of reality and locality as predicted by Bell's celebrated theorem~\cite{bell_einstein_1964,clauser_proposed_1969-1}. One might wonder whether peculiar quantum features could challenge other fundamental notions, like the concept of \emph{causality}, as well. The \emph{process matrix formalism} of Oreshkov, Costa and Brukner~\cite{oreshkov_quantum_2012} was developed to explore this question---studying the most general causal structures compatible with local quantum mechanics for two parties $A$ and $B$.

Surprisingly, the formalism predicts causal structures which are ``causally nonseparable'': they correspond neither to $A$ being before $B$ nor to $B$ being before $A$, nor to a probabilistic mixture thereof. These causal structures can produce correlations incompatible with any definite causal order, violating so-called ``causal inequalities''~\cite{oreshkov_quantum_2012,branciard_simplest_2016}.\footnote{For two parties, the operations implemented to violate causal inequalities have to be quantum~\cite{oreshkov_quantum_2012}; surprisingly, for three or more parties, classical operations are sufficient~\cite{baumeler_maximal_2014}.}

However, the empirical relevance of these results is still completely unclear. Do they appear in some physical situations or are they merely a mathematical artifact of the process matrix formalism?

For three and more parties, there are causally nonseparable processes whose physical realization is known---one instance is the ``quantum switch''~\cite{chiribella_quantum_2013} where the causal order between two parties $A$ and $B$ is controlled by a quantum system belonging to a third party $C$. Processes of this kind, however, \emph{cannot violate} causal inequalities~\cite{araujo_witnessing_2015,oreshkov_causal_2015}: they admit a ``causal model'', i.e., a causally separable process is capable of reproducing their correlations. Their causal nonseparability can only be certified through device-dependent ``causal tomography'' or ``causal witnesses''~\cite{araujo_witnessing_2015}. This is analogous to states which are entangled but cannot violate Bell inequalities, i.e., for which a ``local model'' exists~\cite{werner_quantum_1989}. 

Since the only causally nonseparable processes known to be physically implementable have a causal model, it is tempting to conjecture that the inability to violate causal inequalities without~\cite{brukner_quantum_2014} or with~\cite{oreshkov_causal_2015} operations extended to shared entangled states by all parties \emph{singles out} the physical causal structures from unphysical ones. Ref.~\cite{oreshkov_causal_2015} contains an example of a tripartite process matrix with a causal model but which does not remain causal under extensions, i.e., is not ``extensibly causal'', demonstrating the difference between the two notions for more than two parties. 

In this paper, we provide an example of a bipartite causally nonseparable process with a causal model. Furthermore, we give numerical evidence that \emph{bipartite} nonseparable processes exist which do not violate causal inequalities, even when extended with entanglement. No physical interpretation is known for these processes.

The paper is organized as follows: Sec.~\ref{sec:causal-nonseparability} introduces the process matrix formalism and the definitions of causal nonseparability and causal inequalities. In Sec.~\ref{sec:class-of-processes}, we define a class of two-party causally nonseparable processes and construct a causal model for them. This shows that the sets of causally nonseparable and causal inequality violating processes are distinct also in the bipartite case. Since the causally nonseparable processes with a causal model can be interpreted as the mixture of physically implementable process with an unphysical process, this gives evidence that they are not implementable in nature.

In Sec.~\ref{sec:random}, we provide an algorithm to construct nonseparable processes with a causal model by composing a random causally separable process with a non-completely positive map on one party. Using a random sample generated by a ``hit-and-run'' Markov chain~\cite{smith_monte_1980,smith_efficient_1984}, we also show that nonseparable processes with a causal model have nonzero measure in the space of all processes.

In Sec.~\ref{sec:werner-type}, we construct a family of ``Werner processes'', which mimic the behavior of Werner states~\cite{werner_quantum_1989} with respect to nonlocality and entanglement. The Werner processes' causal nonseparability is more resistant to the introduction of ``white noise'' than its ability to violate causal inequalities. This shows that the analogy between causal nonseparability and causal inequalities on the one hand, and entanglement and Bell inequalities on the other, extends beyond what was previously known~\cite{oreshkov_quantum_2012,brukner_bounding_2015,araujo_witnessing_2015,oreshkov_causal_2015}.

Finally, in Sec.~\ref{sec:extensibly-causal}, we examine the behavior of the processes with added shared entanglement between the parties. While some of the processes we constructed \emph{do violate causal inequalities} when extended in this way (are not ``extensibly causal''), numerical calculations indicate that the ability to violate causal inequalities disappears when adding a little white noise, at which the causal nonseparability is preserved. We conjecture that there are processes which are extensibly causal, and yet not physically implementable.

\section{Causal nonseparability and causal inequalities}
\label{sec:causal-nonseparability}
Quantum circuits can be thought of as a formalization of causal structures with a definite causal order. They consist of \emph{wires}, representing quantum systems, which connect boxes, representing \emph{quantum operations}. While for quantum circuits, the order of the operations is fixed~\cite{chiribella_theoretical_2009}, situations where the order of operations is not well-defined are readily represented in the \emph{process matrix formalism}~\cite{oreshkov_quantum_2012}, which can be thought of as a generalization of the quantum circuit formalism. We will briefly introduce the main elements of the formalism; a more detailed introduction to it can be found in Ref.~\cite{araujo_witnessing_2015}.

A quantum operation maps a density matrix $\rho_{A_I} \in A_I$ to a
density matrix $\rho_{A_O} \in A_O$ (where $A_I$ ($A_O$) denotes the space of linear operators on the Hilbert space $\mathcal H^{A_I}$ ($\mathcal H^{A_O}$)). The most general operations within the quantum formalism are \emph{completely positive} (CP) maps $\mathcal{M}_A: A_I \to A_O$. Using the Choi-Jamio\l kowski~\cite{choi_completely_1975,jamiolkowski_linear_1972} (CJ) isomorphism, one can represent every CP map as an operator acting on the tensor product of the input and output Hilbert spaces:
\begin{equation}
\label{eq:cj}
M_A := \left[\left({\cal I}\otimes{\cal M}_{A} \right)(\ket{I}\rangle\langle\bra{I})\right]^{\mathrm T} \in A_I \otimes A_O,
\end{equation}
where $\mathcal I$ is the identity map and $\ket{I}\rangle:= \sum_{j=1}^{d_{\mathcal{H}_{I}}} \ket{jj} \in \mathcal{H}_{I}\otimes \mathcal{H}_{I}$ is a non-normalized maximally entangled state; $^\text{T}$ denotes matrix transposition in the computational basis. 

\begin{figure}[htp]
  \centering
  \includegraphics{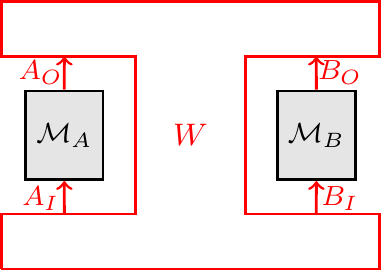}
  \caption{Representation of a bipartite process ${W}$, which linearly maps Alice's and Bob's CP maps $\mathcal M_A, \mathcal M_B$ to a probability. $A_I$ ($B_I$) represents Alice's (Bob's) input Hilbert space and $A_O$ ($B_O$) Alice's (Bob's) output Hilbert space.}
  \label{fig:w-representation}
\end{figure}

The CJ-isomorphism can also be used to represent ``superoperators'' or ``processes'' which map \emph{quantum maps} to quantum maps, quantum states or probabilities~\cite{gutoski_general_2007,chiribella_transforming_2008,chiribella_theoretical_2009,leifer_formulation_2013,oreshkov_quantum_2012}.  In this paper, we will focus on processes mapping \emph{two quantum operations} $\xi_x^a$ and $\eta_y^b$---corresponding to the Choi-Jamio\l{}kowski representation of Alice's and Bob's CP maps---to a \emph{probability} (see Fig.~\ref{fig:w-representation}). Requiring linearity of probabilities in the operations, we can represent it as
\begin{gather}\label{eq:gen-born-rule}
p(\xi_x^a, \eta_y^b) := \tr[W \cdot \xi_x^a \otimes \eta_y^b],\\
W \in A_I \otimes A_O \otimes B_I \otimes B_O.
\end{gather}
To ensure positivity of probabilities for all pairs of possible CP maps (as well as for extended operations where the parties share additional entanglement) the ``process matrix'' $W$ has to be positive semidefinite $W \geq 0$~\cite{oreshkov_quantum_2012}. The normalization of probabilities implies that $\tr[W\cdot \xi^\text{CPTP} \otimes \eta^\text{CPTP}] = 1$ for all CJ-representations of completely positive \emph{trace-preserving maps} $\xi^\text{CPTP}$ and $\eta^\text{CPTP}$~\cite{oreshkov_quantum_2012}. 

We call a process $W^{A \prec B}$ ($W^{B \prec A}$) ``causally ordered'' if it does not allow for signalling from Bob to Alice (Alice to Bob), which is equivalent to the conditions~\cite{araujo_witnessing_2015}
\begin{gather}
W^{A \prec B}  = \tr_{B_O}[W^{A \prec B}] \otimes \id^{B_0}/d_{B_O},\label{eq:valid_W_AB}\\
W^{B \prec A}  = \tr_{A_O}[W^{B \prec A}] \otimes \id^{A_0}/d_{A_O}.\label{eq:valid_W_BB}
\end{gather}

A process matrix $W^\text{sep} \in \mathcal W_\text{sep}$ that can be decomposed into a convex combination ($0\leq q\leq 1$) of causally ordered processes is ``causally separable'':
\begin{equation}
W_{\text{sep}}  =  q  W^{A \prec B}  +  (1{-}q)  W^{B \prec A}.\label{def:caus_sep}
\end{equation}
It was recently shown that one can \emph{efficiently} determine whether a process is causally (non)separable using a \emph{semidefinite program} (SDP). Here we will use the SDP for ``random robustness''~\cite{araujo_witnessing_2015}
\begin{equation}\label{eq:sdp}
\begin{gathered}
\min \lambda \\
\text{s.t.} \quad W = W^{A\prec B} + W^{B\prec A} - \lambda \id^{\circ}, \\
W^{A \prec B}  = \tr_{B_O}[W^{A \prec B}] \otimes \id^{B_0}/d_{B_O},\\
W^{B \prec A}  = \tr_{A_O}[W^{B \prec A}] \otimes \id^{A_0}/d_{A_O},
\end{gathered}
\end{equation}
where $\id^\circ := \id^{A_I A_O B_I B_O}/(d_{A_I} d_{B_I})$. The random robustness $R_r(W)$ is defined as the result of the optimization $R_r(W) := \lambda_\text{opt}$.

If $R_r(W) \le 0$, the SDP gives an explicit decomposition of $W$ into $W^{A \prec B}$ and $W^{B \prec A}$; if $R_r(W) > 0$, the process is not causally separable. The value of $R_r(W)$ is also an operational measure of ``causal nonseparability''. It is related to the minimal amount of ``white noise'' $\id^\circ$ that needs to be mixed with the process to make it causally separable. That is, for $\gamma\geq R_r(W)$, the process $(\gamma\id^\circ + W)/(1+\gamma)$ is causally separable.

A so-called ``dual SDP'' to \eqref{eq:sdp} can then provide the optimal ``causal witness'', i.e., a hermitian operator $S$ such that $\tr [W_{\text{sep}}\, S] \geq 0$ for all causally separable processes $W_\text{sep}$~\cite{araujo_witnessing_2015}. The property $\tr[W\, S] < 0$ can be verified experimentally by measuring a set of operators for Alice and Bob, certifying that the process $W$ is not causally separable. Note that, in analogy to entanglement witnesses~\cite{chruscinski_entanglement_2014}, this certification of causal nonseparability relies on a partial tomography of the process and thus requires trust in Alice's and Bob's local operations: it is ``device-dependent''. 

It is well-known that the entanglement of a quantum state can be certified \emph{device-independently} (without requiring trust Alice's and Bob's operations) if the probability distribution resulting from a set of measurements violates a Bell inequality~\cite{acin_deviceindependent_2007}. In an analogous way, causal nonseparability can be device-independently confirmed using \emph{causal inequalities}~\cite{oreshkov_quantum_2012,branciard_simplest_2016,araujo_witnessing_2015,oreshkov_causal_2015,baumeler_maximal_2014}, where the ``non-causal'' correlations between Alice and Bob alone suffice to show that the process they use is not causally separable, without additional trust in their local operations.

The condition for a probability distribution to be ``causal'', i.e., \emph{not} to violate any causal inequality, is simply that it can be decomposed into a convex combination of a probability distribution which is no-signaling from Bob to Alice\footnote{When Alice is given the input $x$ and outputs $a$ (Bob is given an input $y$ and outputs $b$), no-signaling from Bob to Alice implies that the marginal probability on Alice's side does not depend on Bob's input: $\sum_b p_{A \prec B}(ab|xy) = \sum_b p_{A \prec B}(ab|xy'), \forall y,y'$.} ($p_{A \prec B}$) and a probability distribution which is no-signaling from Alice to Bob ($p_{B \prec A}$)~\cite{branciard_simplest_2016}:
\begin{equation}\label{eq:causal-prob}
p_{\text{causal}} = q p_{A \prec B} + (1-q) p_{B \prec A}.
\end{equation}
Note that the correlations generated by a causally ordered process cannot violate any causal inequality. 

For the scenario where Alice (Bob) has an input bit $x$ ($y$) and outputs one bit $a$ ($b$), one 
causal inequality is a bound on the probability of success of the ``guess your neighbor's input'' (GYNI)~\cite{branciard_simplest_2016}:
\begin{gather}
\label{eq:causal-inequalities}
p_\text{GYNI} := \frac{1}{4}\sum_{x,y} p(a=y,b=x|x,y) \le \frac{1}{2}.
\end{gather}
Some valid processes and local strategies which result in correlations violating \eqref{eq:causal-inequalities} are described in Ref.~\cite{branciard_simplest_2016}.

The relation between causally nonseparable processes and the violation of causal inequalities is not yet fully understood. On the one hand, there exist causally nonseparable processes that can be physically implemented~\cite{procopio_experimental_2015} but have a \emph{causal model}---they do not violate causal inequalities~\cite{araujo_witnessing_2015,oreshkov_causal_2015}. An example of such a process is the ``quantum switch''~\cite{chiribella_quantum_2013}. On the other hand, there are processes that can violate causal inequalities, but it is not known if they can be realized in nature, prompting the conjecture that only processes with a causal model are physically implementable~\cite{brukner_quantum_2014}.

Another natural feature to investigate is the (in)ability for a process to violate causal inequalities, even when extended with an entangled state shared by all parties. We will call processes which do not allow for such a violation ``extensibly causal''~\cite{oreshkov_causal_2015} and come back to this concept in detail in Sec.~\ref{sec:extensibly-causal}.

In the bipartite case, all previously known nonseparable processes violate causal inequalities and it is not clear if nonseparable processes with a causal model even exist~\cite{branciard_simplest_2016}. In the next section, we explicitly provide a class of nonseparable bipartite processes that allow for a causal model. 

\section{Causally nonseparable processes with a causal model}\label{sec:class-of-processes}
We will consider the following class of processes (all the operators are understood to act on qubits, $d_{A_I} = d_{A_O} = d_{B_I} = d_{B_O} = 2$):
\begin{align}
\nonumber
W^{A\prec B}  := & \id^{\circ} + \frac{1}{12} (\id Z Z \id + \id X X \id + \id Y Y \id),\\ \nonumber
W^{B\prec A}  := & \id^{\circ} + \frac{1}{4}(Z \id X Z), \\
W := & q W^{A\prec B} + (1-q+\epsilon) W^{B\prec A} - \epsilon \id^{\circ}. \label{eq:nonseparable-w}
\end{align}
Here $\id, X,Y,Z$ are the Pauli matrices and the tensor products between the Hilbert spaces $A_I, A_O, B_I, B_O$ are implicit, as in the remainder of the paper. The process matrix \eqref{eq:nonseparable-w} is positive semidefinite for $\epsilon \le q-1+ \sqrt{\frac{(1-q)(q+3)}{3}}$ and causally nonseparable for $\epsilon > 0$. As shown in Appendix~\ref{app:nonseparability-w}, its random robustness is $R_r(W) = \epsilon$. It is maximal for $q = \sqrt{3} - 1 \approx 0.732$, where $\epsilon = \frac{4}{\sqrt{3}}-2 \approx 0.309$.

The proof that the process \eqref{eq:nonseparable-w} cannot be used to violate any causal inequalities, for any local strategy\footnote{Note that our proof guarantees the existence of a causal model, which means that the correlations belong to \emph{every} causal polytope, without restriction on the number of inputs and outputs for each party.} consists of two steps: (i)~we show that the set of correlations compatible with $W$ is the same as the set of correlations achievable with $W^{\text{T}_B}$ (where $^{\text{T}_B}$ denotes the partial transpose of the systems $B_I B_O$ with respect to the computational basis); (ii)~we verify that $W^{\text{T}_B}$ is valid and causally separable, hence cannot violate causal inequalities. Taken together, this establishes that $W$ cannot violate any causal inequalities either and therefore admits a causal model.

The first part of the proof is simple. Using definition~\eqref{eq:gen-born-rule} and the self-duality of transposition, we rewrite the probability distribution:
\begin{align}
p(ab|xy) &= \tr[W \xi^a_x \otimes \eta^b_y] = \tr[W^{\text{T}_B} \xi^a_x \otimes (\eta^b_y{})^\text{T}].\label{eq:decomp}
\end{align}
Additionally, for any quantum instrument~\cite{davies_operational_1970} $\{\eta^b_y\}$, the instrument $\{\eta^b_y{}^\text{T}\}$ is also valid, since transposition maps completely positive maps to completely positive maps and trace-preserving maps to trace-preserving maps\footnote{The condition on the CJ representation of a CPTP map is that $\tr_{B_O} M^{B_I B_O} = \id^{B_I}$ and it implies $\tr_{B_O} (M^{B_I B_O}){}^{\text{T}} = \id^{B_I}$.}. This establishes (i), namely that the correlations achievable with $W^{\text{T}_B}$ are the same as those compatible with $W$---note that this holds for \emph{any} process, even when $W^{\text{T}_B}$ is \emph{not positive semidefinite}, and therefore not a valid process matrix. In such a case, the probability distribution will be well-defined for local measurements, but not when extending the process with an entangled state between Alice and Bob, an extension which is physically meaningful and to which we will come back in Sec.~\ref{sec:extensibly-causal}.

For the class of process matrices given in Eq.~\eqref{eq:nonseparable-w}, $W^{\text{T}_B}$ \emph{is} always positive semidefinite. We will now explicitly decompose $W^{\text{T}_B}$ as a convex combination of causally ordered process matrices, proving that it is causally separable (formally, this implies that $R_r(W^{\text{T}_B})\le 0$). 

First, one should notice the similarity of $W^{A\prec B}$ with the process matrix $D^{A\prec B}_{2/3}$ of a depolarizing channel (with $\frac{2}{3}$ probability of depolarizing and $\frac{1}{3}$ probability of perfectly transmitting the state) from Alice to Bob
\begin{equation}\label{eq:process-depol}
D^{A\prec B}_{2/3} := \id^{\circ} + \frac{1}{12}(\id Z Z \id + \id X X \id - \id Y Y \id),
\end{equation}
where only the sign of the term $\id Y Y \id$ differs compared to $W^{A \prec B}$ of \eqref{eq:nonseparable-w}. This exactly corresponds to a partial transpose of the systems $B_I B_O$, such that $W^{A\prec B} = (D^{A\prec B}_{2/3})^{\text{T}_{B}}$.
\begin{figure}[htp]
\includegraphics{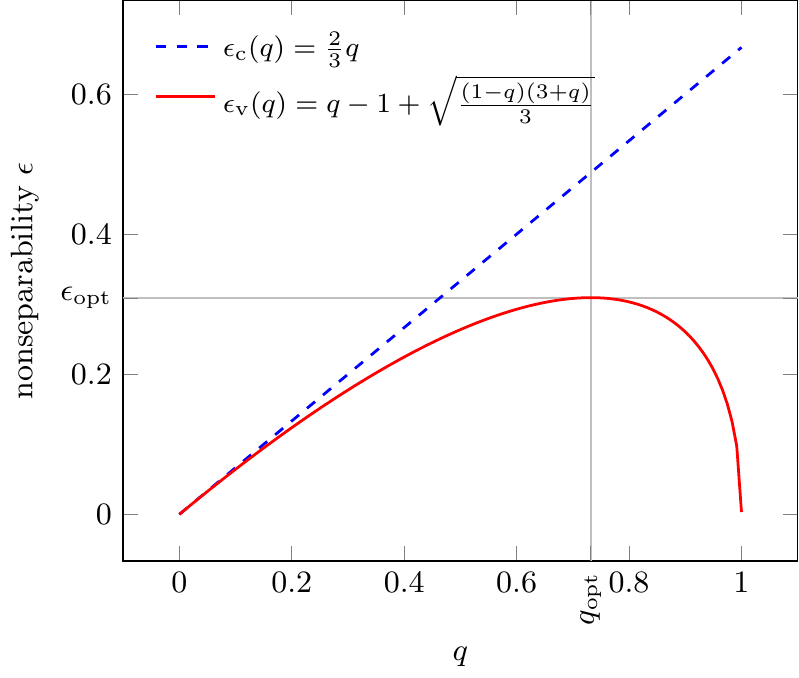}
\caption{For $W$ of Eq.~\eqref{eq:nonseparable-w} to be a valid process, $\epsilon \le \epsilon_{\text{v}}(q)$ (in the region below the red curve). A process only generates causal correlations for $\epsilon \le \epsilon_{\text{c}}(q)$ (the region below the dotted blue curve). Since $\epsilon_{\text{v}}\le \epsilon_{\text{c}}$, every valid process of the form~\eqref{eq:nonseparable-w} allows for a ``causal'' probability distribution. $\epsilon_\text{opt}$ and $q_\text{opt}$ are the parameters maximizing causal nonseparability.}\label{fig:region-non-causality}
\end{figure}
Using the definition of the depolarizing process $D^{A\prec B}_{2/3} = \frac{2}{3} \id^{\circ} + \frac{1}{3} I^{A\prec B}$,
where $I^{A\prec B} = \id^{A_I} \ket{I}\rangle\langle\bra{I}^{A_O B_I}  \id^{B_O}/2$. Since $(W^{B\prec A})^{\text{T}_B} = W^{B\prec A}$, we can write $W^{\text{T}_B}$ as:
\begin{align} \nonumber
W^{\text{T}_B} &= \frac{2q}{3} \id^{\circ} + \frac{q}{3} I^{A\prec B} + (1-q+\epsilon) W^{B\prec A} - \epsilon \id^{\circ} \\ \label{eq:wtb-causally-separable}
&= \frac{q}{3} I^{A\prec B} + (1-q+\epsilon) W^{B\prec A} + \left(\frac{2q}{3} -\epsilon \right) \id^{\circ},
\end{align}
which is a convex decomposition into causally ordered processes as long as $\epsilon \le \epsilon_c(q) =  \frac{2q}{3}$. Since $\epsilon \le q-1+\sqrt{\frac{(1-q)(3+q)}{3}}$ for the process given in Eq.~\eqref{eq:nonseparable-w} to be valid, and $q-1+\sqrt{\frac{(1-q)(3+q)}{3}} \le \epsilon_c(q)$, the whole class of processes defined in Eq.~\eqref{eq:nonseparable-w} cannot violate causal inequalities. For a graphical representation of this relationship, see Fig.~\ref{fig:region-non-causality}.

This concludes the proof and provides an explicit causal model for the process $W$ with the instruments $\{\xi_x^a\}$ and $\{\eta_y^b\}$: the process $W^{\text{T}_B}$ with the instruments $\{\xi_x^a\}$ and $\{\eta_y^b{}^{\text{T}}\}$. Since $W^{\text{T}_B}$ is causally separable, it can be interpreted as a probabilistic mixture of two causally ordered processes. There are infinitely many such decompositions since the term $\id^\circ$ in Eq.~\eqref{eq:wtb-causally-separable} can be split and added to $\frac{q}{3}I^{A\prec B}$ and $(1-q+\epsilon)W^{B\prec A}$ in any proportion. 

$W$ can be taken to be $W^{\text{T}_B}$ composed with a transpose map on $B$. The process matrix $\frac{q}{3}(I^{A\prec B}){}^{\text{T}_B}$ becomes positive (meaning that the associated map is completely positive) when adding at least $\frac{2q}{3}$ of white noise. However, the maximal noise that can be admixed by transferring the $\id^\circ$ term in \eqref{eq:wtb-causally-separable} is $\frac{2q}{3}-\epsilon$, which is strictly smaller. Therefore, the causal model for $W$ suggests a natural interpretation for it as a convex combination of an \emph{unphysical} channel from Alice to Bob (as it is not completely positive) with a \emph{physical} channel from Bob to Alice. This provides some evidence---yet not a proof---that the process $W$ is not physically implementable. 

\section{Random causally nonseparable processes with a causal model}
\label{sec:random}
In this section, we develop a method to construct a broad class of causally nonseparable processes with a causal model. Given a random causally separable process, one applies a positive, but not necessarily completely positive map $Q_B(\cdot)$ on Bob's side:
\begin{equation}\label{eq:rand}
W_\text{sep?} = Q_B(W_\text{sep}).
\end{equation}
If the resulting process has negative eigenvalues, it is discarded; otherwise, it is a valid process matrix. Using the same argument as in the preceding section, we know that the resulting process $W_\text{sep?}$ will have a causal model. Sometimes---and these are the interesting cases---the process will also be causally nonseparable. This can readily be checked this via SDP~\eqref{eq:sdp}.

\begin{figure}[htp]
\includegraphics{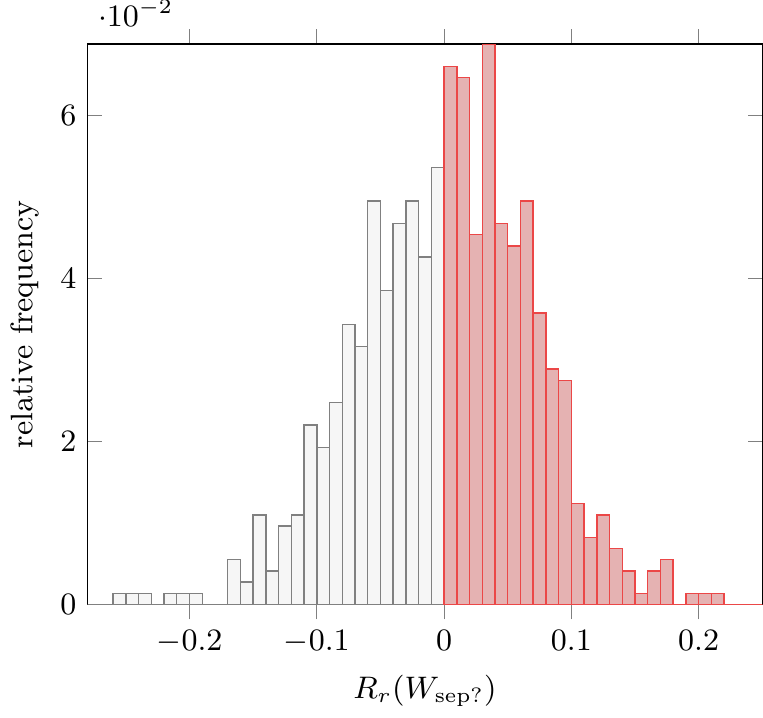}
\caption{Histogram of the random robustness $R_r$ of the subset of $690$ ($\approx 69\%$) valid processes $W_\text{sep?}$ generated from $1000$ uniformly distributed causally separable processes (see Appendix~\ref{app:random}) and applying partial transposition on Bob's side. The $366$ ($\approx 53 \%$) which are causally nonseparable, while admitting a causal model, are represented in red.} \label{fig:rand}
\end{figure}

We generated causally separable process matrices $W_\text{sep}$ (where $d_{A_I} = d_{B_I} = d_{A_O} = d_{B_O} = 2$) according to an asymptotically uniform distribution using the ``hit-and-run'' technique (see Appendix~\ref{app:random} for details) and Eq.~\eqref{eq:rand}, using the transposition map $^{\text{T}_B}$ for $Q_B$. We found that most ($69\%$) of the resulting matrices were positive and hence valid process matrices. About half of these turn out to be causally nonseparable while---by construction---allowing for a causal model (we denote this set by $\mathcal W_\text{nsep}^{\text{(c)}}$). The histogram of the resulting causal nonseparabilities is shown in Fig.~\ref{fig:rand}.
There is therefore a \emph{finite} probability of generating a nonseparable process with a causal model starting from a random of causally separable process. Since the map $^{\text{T}_B}$ is measure-preserving, the set of causally nonseparable processes which admit a causal model is of the same dimension as the set of valid processes $\mathcal W$ itself (see Appendix~\ref{app:measure} for details). 

\section{``Werner'' causally nonseparable processes}
\label{sec:werner-type}
\begin{figure}[htp]
\includegraphics{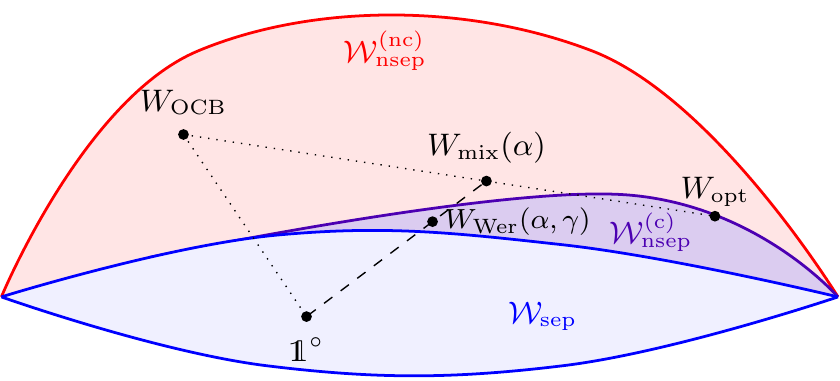}
\caption{Schematic depiction of different types of process matrices. $\mathcal W_\text{nsep}^\text{(nc)}$ is the set of process matrices which can violate causal inequalities and $W_\text{nsep}^\text{(c)}$ is the set of nonseparable process matrices which admit a causal model and $\mathcal W_\text{sep}$ is the set of separable processes. The process $W_\text{mix} :=  \alpha W_\text{opt} + (1-\alpha) W_\text{OCB}$, where $W_\text{opt}$ admits a causal model and $W_\text{OCB}$ doesn't, gives rise to ``Werner type'' processes $W_\text{Wer}(\gamma,\alpha) := (1-\gamma) W_\text{mix} (\alpha) + \gamma \id^\circ$, which have a causal model but are causally nonseparable for a certain level of noise~\eqref{eq:noiselevel}.}\label{fig:sets}
\end{figure}

We will denote the set of causally nonseparable processes as $\mathcal W_\text{nsep}$. It is composed of the set of processes with a causal model ($\mathcal W_\text{nsep}^\text{(c)}$) and the set of processes that can violate causal inequalities ($\mathcal W_\text{nsep}^\text{(nc)}$), which are graphically represented in Fig.~\ref{fig:sets}.

We now construct bipartite processes violating a causal inequality, but which, mixed with some amount of ``white noise'' $\id^\circ$, turn into causally nonseparable processes with a causal model. This behavior is reminiscent of ``Werner states'', which violate Bell inequalities until a noise level of up to $\frac{1}{2}$ but are entangled when mixed with noise up to a level of $\frac{2}{3}$~\cite{werner_quantum_1989}. It shows that the analogy between causal inequalities and causal nonseparability, on the one hand, and Bell inequalities and entanglement, on the other hand, applies to the two-party case.

The idea is to use a convex combination of a process in $\mathcal W_\text{nsep}^\text{(c)}$ and a process in $\mathcal W_\text{nsep}^\text{(nc)}$, which is also invariant under partial transposition with respect to $B$\footnote{Or, in the more general scenario described in the previous section, is invariant under the non-completely-positive operation $Q_B(\cdot)$.}. In this way, one can generate a broad class of ``Werner causally nonseparable processes''.

We will use the process defined in Eq.~\eqref{eq:nonseparable-w} with the maximal causal nonseparability:
\begin{multline}\label{eq:wopt}
W_\text{opt} := (\sqrt{3} - 1) W^{A\prec B} +\frac{1}{\sqrt{3}} W^{B\prec A} \\ - \left(\frac{4}{\sqrt{3}} - 2\right) \id^{\circ},
\end{multline}
together with the process
\begin{equation}\label{eq:wocb}
W_{\text{OCB}} := \id^\circ + \frac{1}{4 \sqrt{2}}(\id Z Z \id + Z \id X Z),
\end{equation}
which was proposed and shown to violate causal inequalities in Ref.~\cite{oreshkov_quantum_2012}. In Appendix~\ref{app:nonseparability-w}, we show that the resulting mixture
\begin{equation}
\label{eq:werner-w}
W_{\text{mix}} (\alpha):= \alpha W_\text{opt} + (1-\alpha) W_\text{OCB}
\end{equation}
has nonseparability $R_\text{mix}(\alpha) := R_r(W_\text{mix}(\alpha)) = \alpha R_r (W_\text{opt}) + (1-\alpha) R_r (W_{\text{OCB}})$, where $R_r (W_{\text{OCB}}) = \sqrt{2}-1$ is the random robustness of $W_\text{OCB}$.

Following the same argument as in the previous section, we can now examine the causal nonseparability of $W_\text{mix}^{\text{T}_B}$, since, by transferring the partial transpose onto Bob's CP maps, we know that it can produce exactly the same correlations as $W_\text{mix}$. Its nonseparability $R_\text{mix}'(\alpha) := R_r(W_\text{mix}^{\text{T}_B}(\alpha))$ is again the weighted average (see Appendix~\ref{app:nonseparability-w}) of the nonseparabilities of $W_\text{opt}^{\text{T}_B}$ and of $W_\text{OCB}^{\text{T}_B} = W_\text{OCB}$:
\begin{align}\nonumber
R_\text{mix}'(\alpha) &=\alpha R_r (W_\text{opt}^{\text{T}_B}) + (1-\alpha) R_r (W_{\text{OCB}})  \\ 
 &=   R_\text{mix}(\alpha) + \frac{2\alpha}{3} R_r (W_\text{opt}^{\text{T}_B}) < R_\text{mix}(\alpha), \label{eq:sep-werner}
\end{align}
where we used $R_r(W_\text{opt}^{\text{T}_B}) = \frac{2 \sqrt{3}-4}{3} < 0$ and $R_r (W_{\text{OCB}}) = \sqrt{2}-1$. This means that there is a finite gap between the nonseparability of $W_\text{mix}$ and the nonseparability of $W_\text{mix}^{\text{T}_B}$.

Using a see-saw algorithm~\cite{branciard_simplest_2016}, we numerically verified that $W_\text{mix}(\alpha)$ indeed violates the causal inequalities of Ref.~\cite{branciard_simplest_2016} as long as $W^{\text{T}_B}_\text{mix}$ is nonseparable ($R_\text{mix}'(\alpha)>0$), i.e., when $\alpha < \frac{3(\sqrt{2}-1)}{1+3\sqrt{2}-2\sqrt{3}} \approx 0.6987$.

This gap translates into a gap between the level of white noise $\id^\circ$ that $W_\text{mix}$ can tolerate before admitting a causal model and the level of noise at which it becomes nonseparable. We therefore define the ``Werner process'' as a  convex combination of $\id^\circ$ and $W_\text{mix}(\alpha)$:
\begin{equation}
W_\text{Wer}(\gamma,\alpha) := (1-\gamma) W_\text{mix} (\alpha) + \gamma \id^\circ.
\end{equation}
Using the definition of nonseparability~\eqref{eq:sdp}, one can verify that the following relations hold (see Appendix~\ref{app:nonseparability-w}):
\begin{equation}
\begin{gathered}
\label{eq:thingsaboutwwer}
W_\text{Wer}\left(\gamma<\frac{R_\text{mix}(\alpha)}{1+R_\text{mix}(\alpha)},\alpha \right) \in \mathcal W_\text{nsep},\\
W_\text{Wer}^{\text{T}_B} \left(\gamma\geq \frac{R_\text{mix}'(\alpha)}{1+R_\text{mix}'(\alpha)},\alpha \right) \in \mathcal W_\text{sep}.
\end{gathered}
\end{equation}
As $W_\text{Wer}$ can violate causal inequalities only if $W_\text{Wer}^{\text{T}_B}$ is causally nonseparable (remember the proof of Sec.~\ref{sec:class-of-processes}), we conclude from \eqref{eq:thingsaboutwwer} that
\begin{equation}
\label{eq:noiselevel}
W_\text{Wer}\left(\frac{R_\text{mix}'(\alpha)}{1+R_\text{mix}'(\alpha)} \leq \gamma<\frac{R_\text{mix}(\alpha)}{1+R_\text{mix}(\alpha)},\alpha \right) \in \mathcal W_\text{nsep}^{\text{(c)}},
\end{equation}
which mimics the behavior of Werner states. See Fig.~\ref{fig:sets} for a graphical representation of the location of $W_\text{Wer}$ with respect to the different sets of processes.

\section{Relationship to extensibly causal processes}
\label{sec:extensibly-causal}
In the context of the physical implementability of process matrices, it is natural to consider the \emph{extension} of a process matrix with an \emph{entangled state shared between the parties}. A process is ``extensibly causal'' if, even when extended with a shared entangled state, it \emph{cannot violate causal inequalities}~\cite{oreshkov_causal_2015}.

Extending a physically implementable process with an entangled state shared among the parties should also result in a physically implementable process. On this account, it is important to consider not only whether a process has a causal model, but rather whether it has such a model when extended with an entangled state. In Ref.~\cite{oreshkov_causal_2015}, an example of a tripartite process with a causal model but which is not extensibly causal was presented, showing that both notions really differ and that the violation of causal inequalities can be ``activated'' by entanglement---and for which no physical implementation is known.

Note that the proof (Sec.~\ref{sec:class-of-processes}) of the existence of causal model for $W$ does not hold when the process is extended with an entangled state between Alice and Bob. It therefore cannot prove that $W$ is extensibly causal. It crucially relies on the fact that the transpose $\{(\eta_y^b)^T\}$ of a valid instrument for Bob $\{\eta_y^b\}$ is still a valid instrument. However, taking the full transpose on Bob's instrument would lead to a ``causal model'' with a partial transpose of the shared entangled state, which can lead to negative probabilities. Conversely, the \emph{partial transpose of Bob's instrument} (with no transposition on Bob's part of the entangled state) is not a valid instrument and does not yield positive probabilities in general.

To numerically study whether $W_\text{opt}$ from Eq.~\eqref{eq:wopt} is extensibly causal, we extended it with a maximally entangled state of two ququarts ($\ket{\phi}^{A_I' B_I'} := \frac{1}{2} (\ket{00} + \ket{11} + \ket{22} + \ket{33})$):
\begin{equation}
\label{eq:w-extended}
W_{\text{ext}} := W_\text{opt} \otimes \ket{\phi}\bra{\phi}^{A_I' B_I'}.
\end{equation}

We chose a maximally entangled ququart state because we believe that extending $W_\text{opt}$ with a higher dimensional state would not improve its ability to violate causal inequalities. 

Using the see-saw algorithm, we optimized $W_\text{ext}$ for a violation of the simplest causal inequalities~\cite{branciard_simplest_2016}. We found that $W_\text{ext}$ \emph{is} able to violate (by about $\SI{8e-5}{}$) the GYNI inequality \eqref{eq:causal-inequalities}, which proves that $W_\text{opt}$ is \emph{not} extensibly causal. Incidentally, it also shows that \emph{in the bipartite case as well}, the violation of causal inequalities can be ``activated'' using entanglement.

\begin{figure}[htp]
\includegraphics{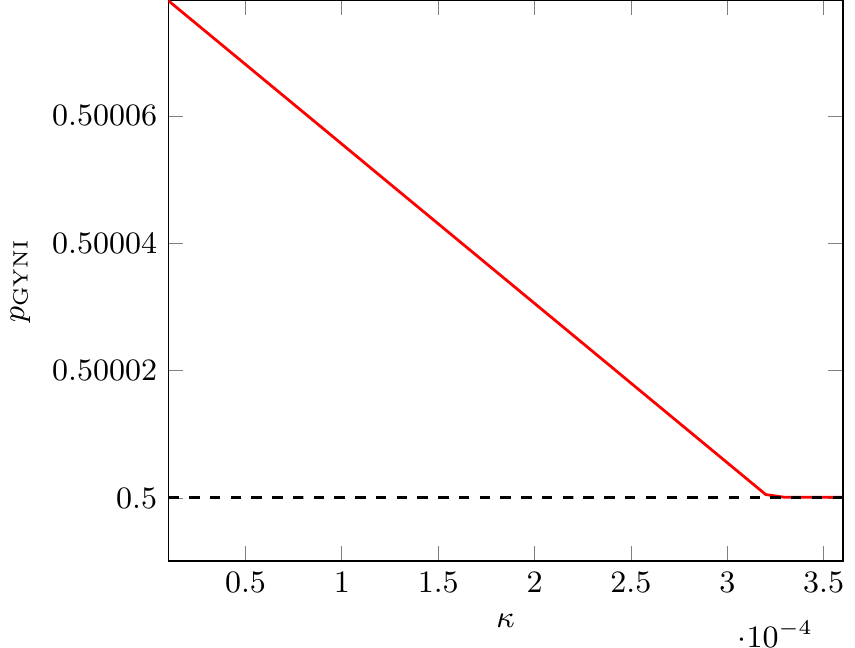}
\caption{Numerically optimized (see-saw) violation of the GYNI inequality Eq.~\eqref{eq:causal-inequalities}, using the noisy extended process $(1-\kappa)W_\text{ext} +\kappa \id^\circ$ (see Eq.~\eqref{eq:w-extended}) and causal bound (dashed). For a noise level of $\kappa > \SI{3.3e-4}{}$, the algorithm fails to find a strategy violating the inequality, as well as for the other known bipartite causal inequalities~\cite{branciard_simplest_2016,oreshkov_quantum_2012}.} \label{fig:see-saw-noise}
\end{figure}

If we adopt the view that extensively causal processes are physical, the activation of violation of causal inequalities, this suffices to exclude $W_\text{opt}$ in the same way as the tripartite process with a causal model but which is not extensibly causal, given in Ref.~\cite{oreshkov_causal_2015}, independently of the argument based on the decomposition of Sec.~\ref{sec:class-of-processes}. However, this is not possible anymore when admixing a small amount of white noise: $(1- \kappa)W_\text{ext} + \kappa \id^\circ$. We ran the see-saw algorithm for different levels of white noise and found no violation of GYNI for $ \kappa>\SI{3.3e-4}{}$ (see Fig.~\ref{fig:see-saw-noise} for a graphical representation of the relationship between noise and violation of GYNI). Similarly, neither the other causal inequality from Ref.~\cite{branciard_simplest_2016} nor the ``original'' causal inequality from Ref.~\cite{oreshkov_quantum_2012} could be violated through see-saw optimization.

\begin{figure}[htp]
\includegraphics{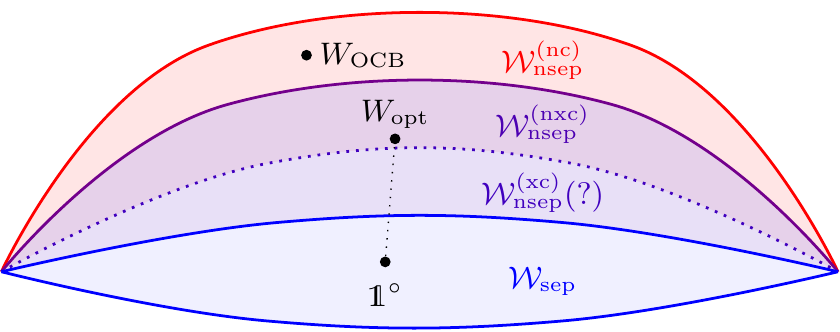}
\caption{Schematic depiction of the sets process matrices with respect to extensible causal separability. $\mathcal W_\text{sep}$ is the set of separable processes, $\mathcal W_\text{nsep}^\text{(nc)}$ is the set of process matrices which can violate causal inequalities, $\mathcal W_\text{nsep}^\text{(nxc)}$ the set of processes with a causal model but which can violate causal inequalities when extended with entanglement, such as $W_\text{opt}$. Based on our numerical evidence (see Fig.~\ref{fig:see-saw-noise}), we conjecture that the set $\mathcal W_\text{nsep}^\text{(xc)}$ of processes which are causally nonseparable and ``extensibly causal'' is not empty. Note that the set of nonseparable processes with a causal model is $\mathcal W_\text{nsep}^\text{(c)} = \mathcal W_\text{nsep}^\text{(xc)} \cup \mathcal W_\text{nsep}^\text{(nxc)}$.}\label{fig:sets2}
\end{figure}

This gives reasonable evidence\footnote{It falls short of being a proof because (i) the entangled state added in $W_\text{ext}$ is finite-dimensional; (ii) the see-saw technique is not guaranteed to converge to the global optimum; (iii) only the three \emph{known} bipartite causal inequalities were tested, inequalities with more settings might still be violated.} that $W_\text{opt}$ mixed with very little white noise is extensibly causal, while still being causally nonseparable (see Fig.~\ref{fig:sets2} for a graphical representation). The argument for unphysicality of Sec.~\ref{sec:class-of-processes} still applies to it. This leads us to conjecture that some extensibly causal processes cannot be physically realized.

\section{Conclusions}
\label{sec:conclusions}
We studied the classification of causally nonseparable process matrices for two parties and found that composing a class of causally separable processes with a transpose map on one party's side results in nonseparable processes with causal models, i.e., that cannot violate causal inequalities.

Since the only interpretation we know relies on applying a non-completely positive map (which is itself is unphysical) to a valid process, the conjecture that processes which do not violate causal inequalities are physically implementable is undermined. 

We also provided a simple algorithm to generate nonseparable processes with causal models---starting from a random separable process and composing it with a positive, but not completely positive map on one party's side. With a finite probability, this yields a nonseparable process with a causal model and shows that the measure of such processes is nonzero within the space of valid processes. The ``hit-and-run'' algorithm we used to generate random process matrices might be of independent interest.

We then developed the analogy between entanglement/nonlocality and causal nonseparability/noncausal correlations by providing a process analogous to a Werner state: it starts having a causal model when mixed with a certain amount of white noise, while still being strictly causally nonseparable.

Finally, we studied whether our processes still have a causal model when extended with an entangled state shared between Alice and Bob (whether they are ``extensibly causal''). The numerical evidence prompted us to conjecture that some of the nonseparable processes we studied are extensibly causal, while not being physically implementable.

An important question remains open: if some processes which have an (extensible) causal model are nonphysical, which other criterion should be used to rule them out? One fairly natural approach is to postulate a ``purification principle'', according to which physically realizable processes can be recovered as part of a pure process in a larger space~\cite{araujo_purification_????}.

\section*{Acknowledgments}
We thank Ognyan Oreshkov and Fabio Costa for pointing out the significance of ``extensible causal separability'' in the context of physical implementability, and Ralph Silva for a useful comment on an earlier version of this manuscript. We acknowledge support from the European Commission project RAQUEL (No. 323970); the Austrian Science Fund (FWF) through the Special Research Programme FoQuS, the Doctoral Programme CoQuS and Individual Project (No. 2462) and the John Templeton Foundation. 

\bibliographystyle{linksen}
\bibliography{physics}
\appendix

\section{Analytic proofs for nonseparabilities in the main text}
\label{app:nonseparability-w}
First, we prove that the class of processes, defined by
\begin{align}
\nonumber
W^{A\prec B}  := & \id^{\circ} + \frac{1}{12} (\id Z Z \id + \id X X \id + \id Y Y \id),\\ \nonumber
W^{B\prec A}  := & \id^{\circ} + \frac{1}{4}(Z \id X Z), \\
W := & q W^{A\prec B} + (1-q+\epsilon) W^{B\prec A} - \epsilon \id^{\circ}, \label{eq:nonseparable-w-app}
\end{align}
has causal separability $R_r(W) = \epsilon$. To do so, we define a causal witness $S_W$ for it:
\begin{equation}
\label{eq:witness}
S_W = \id^\circ - \frac{1}{4}(\id Z Z \id + \id X X \id + \id Y Y \id) - \frac{1}{4} (Z \id X Z).
\end{equation}
We first verify that $S_W$ is a causal witness: $\tr_{A_O} S_W \geq 0$ and $\tr_{B_O} S_W \geq 0$, which is a sufficient condition~\cite{araujo_witnessing_2015} for $S$ to have positive trace with any causally separable process. Therefore, $\tr[W_{\text{sep}} S_W] \geq 0$ and $S_W$ is indeed a causal witness.

We compute $\tr[S_W (W+\lambda \id^\circ)] = -\epsilon + \lambda$, which is negative for $\lambda < \epsilon$ and implies that $R_r(W) \geq \epsilon$. From Eq.~\eqref{eq:nonseparable-w-app}, it is clear that $W+\lambda \id^\circ$ is causally separable for $\lambda \geq \epsilon$, so $R_r(W)\leq \epsilon$. This establishes that $R_r(W) = \epsilon$.

Using the same approach, we can show that the process 
\begin{align}
\label{eq:werner-w-app}
W_{\text{mix}} (\alpha):= \alpha W_\text{opt} + (1-\alpha) W_\text{OCB},
\end{align}
where $W_\text{opt}$ is defined in \eqref{eq:wopt} and $W_\text{OCB}$ in \eqref{eq:wocb}, has nonseparability $R_r (W_\text{mix} (\alpha)) = \alpha R_r(W_\text{opt}) + (1-\alpha) R_r(W_\text{OCB}) =1 + \alpha(\frac{4}{\sqrt{3}}-3)$.

The convexity of random robustness~\cite{araujo_witnessing_2015} implies that the random robustness of a convex combination is smaller than the convex combination of the random robustnesses, so $R_r(W_\text{mix}) \leq 1 + \alpha(\frac{4}{\sqrt{3}}-3)$. Using the same witness $S_W$ as before, we compute $\tr[S_W (W_\text{mix}+\lambda \id^\circ)] = -1 - \alpha(\frac{4}{\sqrt{3}}-3)+\lambda$ which is strictly negative when $\lambda > 1 + \alpha(\frac{4}{\sqrt{3}}-3)$ and implies that $R_r(W_\text{mix}) \geq 1 + \alpha(\frac{4}{\sqrt{3}}-3)$. We conclude that $R_r(W_\text{mix}) =1 + \alpha(\frac{4}{\sqrt{3}}-3)$.

The same proof (with the same witness $S_W$ given in Eq.~\eqref{eq:witness}) can also be used to show that the causal nonseparability of $W_\text{mix}^{\text{T}_B}$ is again the convex combination $R_r(W_\text{mix}^{\text{T}_B}) = \alpha R_r(W_\text{opt}^{\text{T}_B}) + (1-\alpha) R_r(W_\text{OCB})$.

\section{The dimension of the set causally separable processes}
\label{app:measure}
Here we show that the set of causally separable processes $\mathcal W_\text{sep}$ has the same dimension as the set of valid processes $\mathcal W$, which establishes that the set of causally separable processes has nonzero measure in the set of valid processes.

We will use the Hilbert-Schmidt decomposition of operators. An arbitrary process $W$ can be decomposed as $W = \sum_{ijkl=0}^3 \alpha_{ijkl} \sigma_i^{A_I} \otimes \sigma_j^{A_O} \otimes \sigma_k^{B_I} \otimes \sigma_l^{B_O}$.

The condition of normalization of probabilities, i.e., $\tr[W\cdot M_A^\text{CPTP} \otimes M_B^\text{CPTP}] = 1$ for all CJ-representations of completely positive \emph{trace-preserving maps} $M_A^\text{CPTP}$ and $M_B^\text{CPTP}$, implies that some terms of the Hilbert-Schmidt decomposition, corresponding to ``causal loops'' are excluded. In particular, $\alpha_{0j00} = \alpha_{000l} = \alpha_{0j0l} = \alpha_{0jkl} = \alpha_{ij0l} = \alpha_{ijkl} = 0$ for $i,j,k,l\geq 1$ (see the Supplementary Material of Ref.~\cite{oreshkov_quantum_2012}).

Counting all the ``allowed terms'' in the Hilbert-Schmidt decomposition, we find that the dimension $d_W$ of $\mathcal W$ is:
\begin{equation}\label{eq:dim-full-procmat}\nonumber
d_W = (1 + d_{A_I}^2 (d_{A_O}^2-1)) (d_{B_I}^2-1) + (d_{A_I}^2 -1 ) d_{B_I}^2 d_{B_O}^2.
\end{equation}

For causally ordered processes in $W^{A \prec B} \in \mathcal W^{A\prec B}$ compatible with the causal order $A \prec B$, some additional terms, which allow for signaling from Bob to Alice, are excluded in the Hilbert-Schmidt decomposition, reducing the dimension to
\begin{equation}\label{eq:dim-atob}\nonumber
d_{W^{A\prec B}} =  d_{A_I}^2 (1 + (d_{B_I}^2 -1) d_{A_O}^2) -1.
\end{equation}
This means that the set of causally ordered processes $\mathcal W^{A\prec B}$ has measure zero within the set of all process matrices.

Separable processes are convex combinations of $W^{A\prec B}$ and $W^{B\prec A}$. This means that all the terms allowed in the Hilbert-Schmidt decomposition of a valid process matrix are also allowed in the decomposition of separable processes. Therefore $\mathcal W$ and $\mathcal W_\text{sep}$ share the same basis and $d_W = d_{W_\text{sep}}$. 

\section{Generating uniformly distributed processes}
\label{app:random}
We consider the space of process matrices $\mathcal W$ as being embedded in $\mathbb R^{d_W}$. We wish to obtain a \emph{uniform sample} of $\mathcal W$ according to the $d_W$-dimensional volume (Lebesgue measure), which also corresponds to the measure generated by the Hilbert-Schmidt metric. We use an adaptation of the ``hit-and-run'' Markov chain sampler~\cite{smith_monte_1980,smith_efficient_1984} for this task. The iteration works as follows:

\vspace{0.3cm}\textbf{Algorithm 1}\vspace{-0.2cm}
\begin{enumerate}
    \item Select a starting point $W_0$.
    \item Choose a traceless matrix $Q_{i+1}$ from a set of $d_W$ orthogonal traceless matrices and generate a random sign variable $s = \pm 1$.
    \item Find $\mu$ such that $W_i +  \mu  (\id^\circ + s Q_{i+1})$ is on the boundary of the set of valid processes.
    \item Generate a random real scalar $\theta \in [0,\mu]$. Take $W_{i+1} = W_i +  \theta (\id^\circ + s Q_{i+1})$ and go to step 2.
\end{enumerate}

The set of directions is simply the Hilbert-Schmidt basis of allowed terms; it has dimension $d_W$ (see Appendix~\ref{app:measure}). For bipartite processes with $d_{A_I} = d_{B_I} = d_{A_O} = d_{B_O} = 2$, there are $d_W = 87$ possible directions to choose from. 

Finding the intersection with the boundary of the set of positive processes in step 3 turns out to be a semidefinite program
\begin{equation}
\label{eq:sdp-intersection}
\begin{gathered}
\max \mu \\
\text{s.t.} \quad W_i + \mu (\id^\circ + s Q_{i+1}) \geq 0.
\end{gathered}
\end{equation}

However, the SDP which computes $\mu$ at each step of the Markov chain is a bottleneck of the algorithm. Instead, we can skip it and generate $\theta \in [0,1]$, rejecting and retrying if the resulting process is not positive:

\vspace{0.3cm}\textbf{Algorithm 2}\vspace{-0.2cm}
\begin{enumerate}
    \item Select a starting point $W_0$.
    \item Choose a traceless matrix $Q_{i+1}$ from a set of $d_W$ orthogonal traceless matrices and generate a random sign variable $s = \pm 1$.
    \item Generate a random real scalar $\theta \in [0,1]$. Take $W_{i+1} = W_i + \theta (\id^\circ + s Q_{i+1})$.
    \item If $W_{i+1} \geq 0$, go to step 2, otherwise repeat step 3. 
\end{enumerate}

The matrices $\id^\circ + Q_i$ are chosen to be slightly \emph{outside} the set $\mathcal W$ by having slightly negative eigenvalues. Therefore, there is always a finite probability of rejection at step 4, which guarantees that the algorithm samples uniformly all the way up to the boundary. 

The resulting sample $\{W_i\}_{i=0}^{\infty}$ is uniform when two conditions hold~\cite{smith_efficient_1984}. First, from every $W_i, W'$ the probability to have $W_{i+d_W} = W'$ is nonzero, which is indeed true: In $d_W$ steps, one can reach any $W'$ starting from any $W_i$. Second, the uniform distribution is a stationary distribution of the Markov chain. This is also the case: for any $W_i, W'$ the probability to reach $W'$ starting from $W_i$ in $d_W$ steps is the same as the probability to reach $W_i$ starting from $W'$ in $d_W$ steps.

An upper bound on the \emph{convergence} of the hit-and-run algorithm for convex sets (which is the case for the set $\mathcal W$) is known---in particular, the mixing time scales as $\tilde O (d_w^3) := O (d_w^3 \text{polylog}\,d_w)$, which matches the best known mixing times for other algorithms~\cite{lovasz_hitandrun_1999}. For $d_w = 87$ (which is the case for $d_{A_I} = d_{B_I} = d_{A_O} = d_{B_O} = 2$), we would need around $\SI{7e7}{}$ samples to achieve the same statistical significance as from a one-dimensional hit-and-run with $100$ samples, which we deem sufficient for our purposes.

To sample uniformly distributed \emph{causally separable} processes, we use the \emph{rejection method}: after sampling $\SI{7e7}{}$ process matrices (with a warm-up period of $\SI{e6}{}$ discarded steps), we randomly select 1000 causally separable processes (rejecting the $\approx 92.5\%$ of nonseparable ones) in the sample.

\begin{figure}[htp]
\includegraphics{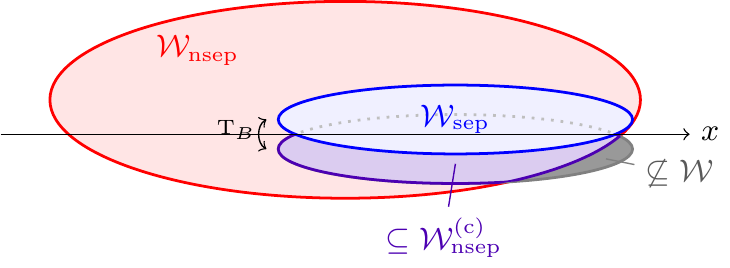}
\caption{Schematic two-dimensional cut of $\mathcal W$ in $\mathrm R^{d_W}$, the partial transpose $^{\text{T}_B}$ here corresponds to a reflection along the horizontal axis $x$. The partial transpose of the set of causally separable processes $\mathcal W_\text{sep}$ consists of three parts: (i)~causally separable matrices ($\subseteq \mathcal W_\text{sep}$), (ii)~non-valid processes ($\not\subseteq \mathcal W$) and (iii)~valid, causally nonseparable matrices with a causal model ($ \subseteq \mathcal W_\text{nsep}^\text{(c)}$).}\label{fig:sets-transpose}
\end{figure}

The map $^{\text{T}_B}$ preserves the Lebesgue measure, since it corresponds to reflections in $\mathbb R^{d_W}$. Therefore, if upon applying $^{\text{T}_B}$ to random causally separable matrices, there is a finite probability to obtain a \emph{valid, causally nonseparable process}, this means that the set of causally nonseparable processes with a separable partial transpose is full dimensional (see Fig.~\ref{fig:sets-transpose}).

\end{document}